\documentclass[manuscript]{aastex}
\usepackage{epsfig}
\usepackage{amsmath}
\begin{document}

\title{Spectroscopic Constants and Vibrational Frequencies for $l$-C$_3$H$^+$
and Isotopologues from Highly-Accurate Quartic Force Fields: The Detection of
$l$-C$_3$H$^+$ in the Horsehead Nebula PDR Questioned}

\date{\today}
\author{Xinchuan Huang}
\affil{SETI Institute, 189 Bernardo Avenue, Suite 100, Mountain View,
California 94043, U.S.A.}
\author{Ryan C. Fortenberry}
\affil{NASA Ames Research Center, Moffett Field, California
94035-1000, U.S.A.}
\author{Timothy J. Lee}
\email{Timothy.J.Lee@nasa.gov}
\affil{NASA Ames Research Center, Moffett Field, California
94035-1000, U.S.A.}

\begin{abstract}

Very recently, molecular rotational transitions observed in the
photon-dominated region of the Horsehead nebula have been attributed to
$l$-C$_3$H$^+$.  In an effort to corroborate this finding, we employed
state-of-the art and proven high-accuracy quantum chemical techniques to
compute spectroscopic constants for this cation and its isotopologues.  Even
though the $B$ rotational constant from the fit of the observed spectrum and
our computations agree to within 20 MHz, a typical level of accuracy, the $D$
rotational constant differs by more than 40\%, while the $H$ rotational
constant differs by three orders of magnitude.  With the likely errors in the
rotational transition energies resulting from this difference in $D$ on the
order of 1 MHz for the lowest observed transition ($J = 4 \rightarrow 3$) and
growing as $J$ increases, the assignment of the observed rotational lines from
the Horsehead nebula to $l$-C$_3$H$^+$ is questionable.

\end{abstract}

\maketitle


\section{Introduction}

Within the past months, $l$-C$_3$H$^+$ has been attributed as the source of
several rotational lines observed in the Horsehead nebula photon-dominated
region (PDR) \citep{Pety12}.  In order for the observed spectral features to be
linked to a single molecular carrier, the molecule must be linear with a
$^1\Sigma$ ground state and a lone ($B$-type) rotational constant close to
11.24 GHz.  However, only standard quantum chemical computations of
$l$-C$_3$H$^+$ have given qualitative corroborating evidence for the rotational
constants, and no quantitative, high-accuracy reference data, either from
computation or laboratory experiment, exists for $l$-C$_3$H$^+$.  The
corresponding C$_3$H radical has long been known to exist in the interstellar
medium (ISM) \citep{Thaddeus85} and was confirmed with concurrently generated
laboratory reference data \citep{Gottlieb85}.  Although laboratory studies of
known interstellar molecules in simulated interstellar environments have often
produced what is believed to be C$_3$H$^+$ \citep{Boehme83, Schwell12} with
theory providing further insights \citep{Radom76, Ikuta96, Wang07}, neither
high-accuracy rotational nor vibrational spectroscopic data has yet been
generated for $l$-C$_3$H$^+$.


Recently, we have employed state-of-the art $ab$ $initio$ quartic force fields
(QFFs) and perturbation theory at second order \citep{Papousek82} to compute
rotational constants that are within 20 MHz or better of experiment
\citep{Fortenberry11HOCO, Fortenberry11cHOCO, Fortenberry12hococat,
Fortenberry12HOCScat} for similar tetra-atomic systems.  The use of QFFs has
also been applied to other molecules of interstellar relevance in order to
compute accurately their spectroscopic constants in addition to their
fundamental vibrational frequencies and even some overtones and combination
bands \citep{Huang08, Lee09, Huang09, Inostroza11, Huang11, Huang11D,
Fortenberry12HOCScat, Fortenberry13HCN}.  Since high-accuracy reference data is
currently not available for $l$-C$_3$H$^+$, the methods utilized previously are
employed here in order to assist in confirmation for the detection of this
cation in the Horsehead nebula.  Additionally, we will also supply
spectroscopic constants and vibrational frequencies for the deuterated form and
for each of the three $^{13}$C singly-substituted isotopologues.

\section{Computational Details}

Utilizing the MOLPRO 2006.1 suite of quantum chemical programs \citep{MOLPRO},
spin-restricted Hartree-Fock (RHF) \citep{ScheinerRHF87} coupled cluster theory
\citep{Lee95Accu, Shavitt09, ccreview} at the singles, doubles, and
perturbative triples level [CCSD(T)] \citep{Rag89} is employed to compute the
reference geometry.  This initial geometry is based on a composite approach to
give the most accurate bond lengths in this linear molecule.  The CCSD(T) bond
lengths (denoted as $R$) computed with the cc-pV5Z basis set \citep{Dunning89,
cc-pVXZ, Dunning01} are corrected for effects from the smaller cc-pVQZ basis
set as well as core correlation and scalar relativistic effects
\citep{Douglas74} through the following approach:
\begin{equation}
R=R_{5Z} + (R_{5Z} - R_{QZ}) + (R_{Rel.+Core} - R_{QZ}).
\label{bondlengths}
\end{equation}
The $R_{Rel.+Core}$ CCSD(T) geometry optimization makes use of the standard
cc-pVQZ-DK basis set, which can account for relativistic effects, further
augmented to include core correlating $s$, $p$, $d$, and $f$ functions from the
standard cc-pCVQZ basis set \citep{Dunning89}.  The last term in
Eq.~\ref{bondlengths} serves to isolate the scalar relativistic and core
correlating effects so that they may then enhance the standard cc-pV5Z bond
lengths.

From this reference geometry, the quartic force field (QFF) is created from
displacements of 0.005 \AA\ and 0.005 radians \citep{Huang11}, respectively for
the three bond lengths and two doubly-degenerate linear bends, and contains 569
points.  At each point CCSD(T)/cc-pVTZ, cc-pVQZ, and cc-pV5Z energies are
computed and extrapolated to the complete basis set (CBS) limit via a
three-point formula \citep{Martin96}.  The CBS energy is then corrected for
scalar relativity \citep{Douglas74} and core correlation effects from the
Martin-Taylor basis set \citep{Martin94}.  This results in the CcCR QFF defined
previously \citep{Fortenberry11HOCO, Fortenberry13CH2CN-}.  The QFF is then fit
with a sum of residual squares of $5.7 \times 10^{-17}$ a.u.$^2$ to a
Taylor series expansion.  Since the reference geometry is close to but not identical to the exact minimum, the QFF is then transformed to the exact minimum leading to a potential of the form:
\begin{equation}
V=\frac{1}{2}\sum_{ij}F_{ij}\Delta_i\Delta_j +
\frac{1}{6}\sum_{ijk}F_{ikj}\Delta_i\Delta_j\Delta_k +
\frac{1}{24}\sum_{ijkl}F_{ikjl}\Delta_i\Delta_j\Delta_k\Delta_l
\label{VVib}
\end{equation}
where $F_{ij\ldots}$ are force constants and $\Delta_i$ are the displacements.
For further details on this procedure see \cite{Huang08}.

This high-symmetry $C_{\infty v}$ system is most easily represented in the
simple-internal coordinate system defined here (with atom numbering from
Fig.~\ref{fig}) as:
\begin{align}
S_1(\Sigma^+) &=\mathrm{C_1}-\mathrm{H}\\
S_2(\Sigma^+) &=\mathrm{C_1}-\mathrm{C_2}\\
S_3(\Sigma^+) &=\mathrm{C_2}-\mathrm{C_3}\\
S_4(\Pi_{xz}) &=\mathrm{LIN1}(\mathrm{H}-\mathrm{C_1}-\mathrm{C_2}-\vec{y})\\
S_5(\Pi_{yz}) &=\mathrm{LIN1}(\mathrm{H}-\mathrm{C_1}-\mathrm{C_2}-\vec{x})\\
S_6(\Pi_{xz}) &=\mathrm{LIN1}(\mathrm{C_1}-\mathrm{C_2}-\mathrm{C_3}-\vec{y})\\
S_7(\Pi_{yz}) &=\mathrm{LIN1}(\mathrm{C_1}-\mathrm{C_2}-\mathrm{C_3}-\vec{x})
\end{align}
These coordinate subscripts correspond to the force constants given in Table
\ref{fc}.  The $\vec{x}$ and $\vec{y}$ quantities are necessary to define the
linear bending angles ($S_4-S_7$) and correspond to a direction perpendicular
to the bending plane so that degenerate linear modes may be uniquely specified,
since $\vec{x} \perp \vec{y}$.  Second-order vibrational perturbation theory
(VPT2) \citep{Mills72, Watson77} via the SPECTRO program \citep{spectro}
produces the desired spectroscopic constants and vibrational frequencies.
Since the QFF is computed within the Born-Oppenheimer approximation, only
the SPECTRO anharmonic analyses differ between the five total isotopologues.

\section{Discussion}


The $l$-C$_3$H$^+$ and isotopologue bond lengths, rotational constants, quartic
and sextic centrifugal distortion constants, and center-of-mass dipole moments
are given in Table \ref{StructHarm} for this $^1 \Sigma^+$ cation. The
fundamental vibrational frequencies are listed in Table \ref{vptvci}.
Exploratory CCSD(T)/cc-pVTZ computations indicate that a $^3 \Sigma^+$ state
exists 40.3 kcal/mol (1.75 eV) higher than the ground $^1 \Sigma^+$ state, and
the $^3 \Sigma^+$ C$_3$H$^+$ $B_e$ is more than 350 MHz higher than the $^1
\Sigma^+$ C$_3$H$^+$ $B_e$.  Hence, effects from the $^3 \Sigma^+$ state should
not be important.

The relevant $^1 \Sigma^+$ C$_3$H$^+$ data are in fairly good agreement with
the full-dimensional PES computed previously given the much lower level of
theory used \citep{Wang07}.  The inclusion of the scalar relativity correction
is necessary since, for example, $B_0$ for standard $l$-C$_3$H$^+$ computed
with only the CBS-energy-including QFF is 11 209.78 MHz, while the CR QFF
(which additionally includes the relativistic terms, see
\cite{Fortenberry11HOCO}) raises this value to 11 213.99 MHz.  Similarly but
more importantly, core correlation is also necessary as the rotational constant
for the CcC QFF, which is the CBS plus core correlation QFF, puts $B_0$ at 11
258.41 MHz.  The total CcCR QFF $B_0$, listed in Table \ref{StructHarm}, is
thus 11 262.68 MHz.  This value differs by 17.73 MHz from that (11 244.947 4
MHz) obtained in the fitting by \cite{Pety12} to assign the observed lines in
the Horsehead nebula to $l$-C$_3$H$^+$.  This represents similar accuracy as
that noted previously for the $cis$- and $trans$-HOCO radicals
\citep{Fortenberry11HOCO, Fortenberry11cHOCO}, $trans$-HOCO$^+$
\citep{Fortenberry12hococat}, and $trans$-HOCS$^+$
\citep{Fortenberry12HOCScat}, as well as many other examples.

However, the correspondence between theory and experiment for the $D$-type
constant is not as good.  The CcCR $D_e$ value is 4.248 kHz while the $D$ value
obtained by \cite{Pety12} in their second-order fit is 7.652 kHz and 7.766 kHz
for their third-order fit.  This represents a difference of 3.404 kHz or 44.5\%
between theory and experiment for the second-order fit $D$ and is larger still
for the third-order fit.  Though this is a comparison of $D_e$ to $D_0$, i.e.,
vibrational averaging is not included in the theoretical value, the percentage
change between the computed $B_0$ and $B_e$ values for this same molecule is
only 0.77\%.  Further, various QFFs have been formulated to examine this system
(33 in all) ranging from those that are state-of-the art and have been proven
to yield highly-accurate spectroscopic constants including $D$, such as the
given CcCR QFF, to those that use more modest levels of theory, such as
CCSD(T)/cc-pVTZ, and the range in $D_e$ values for these QFFs is only 0.053
kHz, which conclusively shows that the $ab$ $initio$ $D_e$ value is converged
to an extent that rules out a value near 7.6-7.8 kHz.

Additionally, other linear systems have been studied using similar theoretical
approaches.  For instance, the percentage error between the computed $D_e$
(84.135 kHz) and the experimental $D$ [87.22 kHz from \cite{Winnewisser71}] is
3.5\% for HCN computed with a CCSD(T)/ANO1 QFF \citep{Lee93HCN}.  The $D$/$D_e$
error for CCH$^-$ is 2.1\% as computed with a CcCRE  QFF (with corrections for
higher-order electron correlation effects: ``E") in the same second order
perturbational approach utilized here, where $D_e$ is 94.3 kHz \citep{Huang09},
and the experimental $D$ is 96.97 kHz \citep{Brunken07}. Furthermore, the error
for acetylene is only 2.27\% between a $D_e$ of 47.673 kHz \citep{Martin98} and
a $D$ of 48.780 kHz \citep{Kabbadj91}.  Perhaps a more relevant example for the
current case is given for C$_3$N$^-$ where $D_0$ has been experimentally
determined to be 0.68 578 kHZ and a CCSD(T)/aug-cc-pVQZ QFF yields a $D_e$ of
0.627 kHz \citep{Thaddeus08C3N-, McCarthy08C3N-}.  The 8.57\% error is still
much less than the 44.5\% discrepancy for C$_3$H$^+$ and is probably somewhat
larger then the other examples because of using lower levels of theory to
compute the QFF, and because it is well known that small molecular anions are
harder to describe accurately.  Hence, the known errors for these systems are
an order of magnitude smaller than that which is present for $D_e$ and $D$ in
$l$-C$_3$H$^+$.  As a result, we cannot attribute the substantial disagreement
between the computed $D_e$ for $l$-C$_3$H$^+$ and the $D$ value obtained by
\cite{Pety12} merely to errors in the theoretical approach.  Thus, it is
doubtful that the series of lines observed in the Horsehead nebula actually
correspond to $l$-C$_3$H$^+$.

If the computed $D_e$ value closely represents the actual $D$ for
$l$-C$_3$H$^+$ and assuming the $B_0$ value reported by \cite{Pety12}, the $J =
4 \rightarrow 3$ transition will differ from that reported by \cite{Pety12} by
0.9 MHz.  The $J = 5 \rightarrow 4$ transition will differ by 1.7 MHz, $J = 6
\rightarrow 5$ by about 3.0 MHz, and these differences will continue to
increase up to 23.4 MHz as $J$ increases to the $J = 12 \rightarrow 11$
transition.  For high-resolution rotational spectroscopy, these large
discrepancies are well outside of the precision present in most spectrometers
utilized in the laboratory and for observation of the ISM, and well outside the
errors expected from state-of-the art $ab$ $initio$ calculations.  For example,
\cite{Brites12} computed the rotational transitions of HCNH$^+$ with a
six-dimensional PES from a CCSD(T)-F12A/cc-pCVTZ-F12 QFF for $J = 1 \rightarrow
0$ up to $J = 10 \rightarrow 9$ with errors ranging from 0.6 MHz to 4.84 MHz
compared to experimental work \citep{Amano06}. This comparison of theory and
experiment again strongly supports our assertion that the assignment of the
observations in the Horsehead Nebula PDR to $l$-C$_3$H$^+$ are incorrect.

Similarly, the computed $H_e$ value for $l$-C$_3$H$^+$ is 0.375 mHz, while the
third-order fit $H$ obtained by \cite{Pety12} is 0.56 Hz or 560 mHz.  This
represents three orders of magnitude difference between the two values.  There
are not as many comparisons in the literature between theoretical $H_e$ values
and experimental $H$ values, but for acetylene, $H_e$ has been computed to be
0.0380 Hz \citep{Martin98}, while $H$ has been experimentally determined to be
0.0480 Hz \citep{Kabbadj91}, an error of 20.8\%.  Even if the computed $H_e$
for $l$-C$_3$H$^+$ is in error from the proper physical value by as much as
40\%, this is only a minute step towards the value necessary to fit the lines
observed.  Admittedly, comparison of computed $H_e$ values to experimental $H$
constants has not been as common making the error range for the computed value
less certain, but the substantially large difference between values here should
be well-beyond the potential accuracy range for the computed $H_e$.  This
difference in the $H$ constant casts further doubt on $l$-C$_3$H$^+$ as the
carrier of the observed transitions in the Horsehead nebula PDR.


Since the energy points necessary to define the QFF are computed within the
Born-Oppenheimer approximation, the same QFF can also be used to produce VPT2
spectroscopic constants and vibrational frequencies for the isotopologues, as
well.  Thus, the spectroscopic constants of C$_3$D$^+$, $^{13}$CCCH$^+$,
C$^{13}$CCH$^+$, and CC$^{13}$CH$^+$ are listed in Table \ref{StructHarm}, and
the fundamental vibrational frequencies are given in Table \ref{vptvci}.  These
reference data should aid in the analysis of further laboratory studies of
interstellar reactions that create $l$-C$_3$H$^+$.

\section{Conclusions}

Even though the CcCR VPT2 computed $B_0$ rotational constant for $l$-C$_3$H$^+$
differs by less than 20 MHz from that obtained by \cite{Pety12} to assign lines
in the Horsehead nebula PDR to this cation, the more substantial differences
between the $D$ and $H$ constants strongly questions the assignment of
$l$-C$_3$H$^+$ to the observed lines.  These large discrepancies, especially
for the $D$ constant, will alter the rotational spectrum of $\tilde{X}\
^1\Sigma^+$ C$_3$H$^+$ by nearly 1 MHz for the lowest observed rotational
transition, $J = 4 \rightarrow 3$, and by more than 23 MHz for the highest, $J
= 12 \rightarrow 11$.  Hence, the astronomical lines observed cannot be
conclusively linked to this cation as the carrier.

It is hoped that the reference data provided here should assist in laboratory
studies that can conclusively determine the rotational spectrum of
$l$-C$_3$H$^+$.  Additionally, the data are extended to the deuterated and
$^{13}$C singly-substituted isotopologues as well.  C$_3$H$^+$ has long been
hypothesized to exist in the ISM and is of significance in the carbon chemistry
of space \citep{Turner00, Wakelam10}.  There is little doubt that it should be
present in ISM, but the attribution of the observed lines by \cite{Pety12} to
$l$-C$_3$H$^+$ is questionable.

\section{Acknowledgements}

XH received funding from the NASA/SETI Institute Cooperative Agreements
NNX12AG96A.  The NASA Postdoctoral Program administered by Oak Ridge Associated
Universities funded RCF.  Support from NASA's Laboratory Astrophysics `Carbon
in the Galaxy' Consortium Grant (NNH10ZDA001N) is gratefully acknowledged.  The
CheMVP program was used to created Fig.~\ref{fig}.  Additionally, the authors
thank Dr. Michael C. McCarthy of the Harvard-Smithsonian Center for
Astrophysics and Dr. David W. Schwenke of NASA Ames Research Center for useful
discussions and explorations related to this project.

\bibliographystyle{apj}

\newpage

\begin{figure}[h]
\caption{The CcCR equilibrium geometry of $\tilde{X}\ ^1\Sigma^+$ C$_3$H$^+$.}
\label{fig}
\end{figure}

\renewcommand{\baselinestretch}{1}
\begingroup
\begin{table}[h]

\caption{The simple-internal C$_3$H$^+$ CcCR QFF Force Constants (in
mdyn/\AA$^n$$\cdot$rad$^m$).}

\label{fc}

\centering
\small

\begin{tabular}{c r c r c r c r c r}
\hline

F$_{11}$ & 5.810 273 & F$_{441}$ & -0.1688 & F$_{2222}$ &405.81 & F$_{5532}$ & -0.16 & F$_{7511}$ &  0.28 \\
F$_{21}$ &-0.214 189 & F$_{442}$ & -0.6565 & F$_{3111}$ &  0.05 & F$_{5533}$ &  0.22 & F$_{7521}$ &  0.94 \\
F$_{22}$ &13.840 846 & F$_{443}$ & -0.1210 & F$_{3211}$ &  0.22 & F$_{5544}$$^a$ &  0.41 & F$_{7522}$ & -0.11 \\
F$_{31}$ & 0.075 121 & F$_{551}$ & -0.1688 & F$_{3221}$ &  0.22 & F$_{5555}$ &  1.22 & F$_{7531}$ &  0.01 \\
F$_{32}$ & 0.116 818 & F$_{552}$ & -0.6565 & F$_{3222}$ &  2.09 & F$_{6411}$ &  0.28 & F$_{7532}$ &  0.36 \\
F$_{33}$ & 8.520 119 & F$_{553}$ & -0.1210 & F$_{3311}$ &  0.15 & F$_{6421}$ &  0.94 & F$_{7533}$ & -0.35 \\
F$_{44}$ & 0.330 792 & F$_{641}$ & -0.0004 & F$_{3321}$ &  0.30 & F$_{6422}$ & -0.11 & F$_{7544}$$^a$ & -0.04 \\
F$_{55}$ & 0.330 792 & F$_{642}$ &  0.0412 & F$_{3322}$ & -1.51 & F$_{6431}$ &  0.01 & F$_{7555}$ & -0.11 \\
F$_{64}$ & 0.004 472 & F$_{643}$ &  0.1390 & F$_{3331}$ & -0.05 & F$_{6432}$ &  0.36 & F$_{7654}$$^a$ & -0.05 \\
F$_{66}$ & 0.034 854 & F$_{661}$ & -0.0212 & F$_{3332}$ & -0.98 & F$_{6433}$ & -0.35 & F$_{7665}$$^a$ & -0.03 \\
F$_{75}$ & 0.004 472 & F$_{662}$ & -0.2630 & F$_{3333}$ &245.44 & F$_{6444}$ & -0.11 & F$_{7711}$ & -0.08 \\
F$_{77}$ & 0.034 854 & F$_{663}$ &  0.0347 & F$_{4411}$ & -0.08 & F$_{6554}$$^a$ & -0.04 & F$_{7721}$ &  0.10 \\
F$_{111}$ & -33.4159 & F$_{751}$ & -0.0004 & F$_{4421}$ &  0.45 & F$_{6611}$ & -0.08 & F$_{7722}$ &  0.18 \\
F$_{211}$ &   0.1250 & F$_{752}$ &  0.0412 & F$_{4422}$ & -0.18 & F$_{6621}$ &  0.10 & F$_{7731}$ &  0.11 \\
F$_{221}$ &   0.0463 & F$_{753}$ &  0.1390 & F$_{4431}$ & -0.22 & F$_{6622}$ &  0.18 & F$_{7732}$ & -0.38 \\
F$_{222}$ & -82.4686 & F$_{771}$ & -0.0212 & F$_{4432}$ & -0.16 & F$_{6631}$ &  0.11 & F$_{7733}$ & -0.19 \\
F$_{311}$ &  -0.0121 & F$_{772}$ & -0.2630 & F$_{4433}$ &  0.22 & F$_{6632}$ & -0.38 & F$_{7744}$ &  0.00 \\
F$_{321}$ &  -0.0241 & F$_{773}$ &  0.0347 & F$_{4444}$ &  1.22 & F$_{6633}$ & -0.19 & F$_{7755}$ & -0.10 \\
F$_{322}$ &  -0.7100 & F$_{1111}$ & 172.62 & F$_{5511}$ & -0.08 & F$_{6644}$ & -0.10 & F$_{7764}$$^a$ & -0.03 \\
F$_{331}$ &  -0.1360 & F$_{2111}$ &   0.14 & F$_{5521}$ &  0.45 & F$_{6655}$ &  0.00 & F$_{7766}$$^a$ &  0.02 \\
F$_{332}$ &  -0.4027 & F$_{2211}$ &  -0.69 & F$_{5522}$ & -0.18 & F$_{6664}$ & -0.10 & F$_{7775}$ & -0.10 \\
F$_{333}$ & -49.4860 & F$_{2221}$ &   0.98 & F$_{5531}$ & -0.22 & F$_{6666}$ &  0.06 & F$_{7777}$ &  0.06 \\
 
\hline
\end{tabular}

$^a$These necessary force constants are not symmetry-unique but are defined
from relationships of other force constants.  See Ref.~60 of \cite{Martin98}.\\
 
\end{table}
\endgroup
\renewcommand{\baselinestretch}{2}

\renewcommand{\baselinestretch}{1}
\begingroup
\begin{table}[h]

\caption{The CcCR QFF Zero-Point ($R_{\alpha}$ vibrationally-averaged) and
Equilibrium Structures, Rotational Constants, CCSD(T)/aug-cc-pV5Z Dipole
Moment, Vibration-Rotation Interaction Constants, and Quartic and Sextic
Distortion Constants of linear $\tilde{X}\ ^1\Sigma^+$ C$_3$H$^+$, C$_3$D$^+$,
$^{13}$CCCH$^+$, C$^{13}$CCH$^+$, and CC$^{13}$CH$^+$.}

\label{StructHarm}

\centering

\scriptsize

\begin{tabular}{l | r r r r r} 
\hline\hline

                   & C$_3$H$^+$    & C$_3$D$^+$    & $^{13}$CCCH$^+$& C$^{13}$CCH$^+$& CC$^{13}$CH$^+$ \\
\hline
r$_0$(C$_1-$H)     & 1.069 951 \AA & 1.072 922 \AA & 1.067 560 \AA  & 1.070 095 \AA  & 1.069 996 \AA \\
r$_0$(C$_1-$C$_2$) & 1.230 766 \AA & 1.231 576 \AA & 1.230 543 \AA  & 1.230 961 \AA  & 1.230 903 \AA \\
r$_0$(C$_2-$C$_3$) & 1.332 581 \AA & 1.331 723 \AA & 1.332 879 \AA  & 1.332 800 \AA  & 1.332 416 \AA \\
$B_0$              & 11 262.68 MHz & 10 115.99 MHz & 10 824.96 MHz  & 11 258.19 MHz  & 10 919.88 MHz \\
$\alpha^B$ 1       &  33.5 MHz     &  49.0 MHz     &  31.5 MHz      &  33.2 MHz      &  30.6 MHz \\
$\alpha^B$ 2       &  74.8 MHz     &  55.0 MHz     &  71.5 MHz      &  72.2 MHz      &  73.4 MHz \\
$\alpha^B$ 3       &  37.1 MHz     &  30.1 MHz     &  35.4 MHz      &  37.2 MHz      &  35.1 MHz \\
$\alpha^B$ 4       &  -2.6 MHz     & -10.2 MHz     &  -2.3 MHz      &  -2.5 MHz      &  -1.7 MHz \\
$\alpha^B$ 5       &-157.3 MHz     &-133.2 MHz     &-151.2 MHz      &-157.7 MHz      & 152.7 MHz \\
$\tau_{aaaa}$$^a$  & -16.992 kHz   & -12.991 kHz   & -15.803 kHz    & -16.988 kHz    & -15.966 kHz \\
\hline                                                    
r$_e$(C$_1-$H)$^b$ & 1.078 961 \AA & --            & --             & --             & --\\
r$_e$(C$_1-$C$_2$) & 1.235 360 \AA & --            & --             & --             & --\\
r$_e$(C$_2-$C$_3$) & 1.339 841 \AA & --            & --             & --             & --\\
$B_e$              & 11 175.51 MHz & 10 039.77 MHz & 10 740.68 MHz  & 11 175.30 MHz  & 10 834.97 MHz\\
$D_e$              & 4.248 kHz     & 3.248 kHz     & 3.951 kHz      & 4.247 kHz      & 3.992 kHz\\
$H_e$$^c$          & 0.375 mHz     & 0.315 mHz     & 0.315 mHz      & 0.375 mHz      & 0.361 mHz\\
$\mu_z$$^d$        & 3.06 D        & --            & --             & --             & --\\ 

\hline
\end{tabular}

$^a$Since this is a linear molecule, $\tau_{aaaa}=\tau_{bbbb}=\tau_{aabb}$ while 
all other quartic centrifugal distortion constants are 0.0.\\
$^b$The use of the Born-Oppenheimer approximation necessitates that the
equilibrium geometry be the same for each isotopologue.\\
$^c$$H_e=\Phi_{aaa}=\Phi_{bbb}$, and all other sextic centrifugal distortion
constants are 0.0.\\
$^d$The C$_3$H$^+$ coordinates (in \AA\ with the center-of-mass at the origin)
used to generate the Born-Oppenheimer dipole moment component are: H, 0.000000,
0.000000, -2.285175; C$_1$, 0.000000, 0.000000, -1.206214; C$_2$, 0.000000,
0.000000, 0.029147; C$_3$, 0.000000, 0.000000, 1.368989.\\

\end{table}
\endgroup

\begingroup
\begin{table}[h]

\caption{The VPT2 CcCR QFF harmonic and anharmonic fundamental vibrational
frequencies (in cm$^{-1}$) for C$_3$H$^+$, C$_3$D$^+$, $^{13}$CCCH$^+$,
C$^{13}$CCH$^+$, and CC$^{13}$CH$^+$.}

\centering
\scriptsize
\begin{tabular}{c l | c c | c c | c c | c c | c c}
\hline\hline
\label{vptvci}

 & & \multicolumn{2}{c|}{C$_3$H$^+$} & \multicolumn{2}{c|}{C$_3$D$^+$} &
\multicolumn{2}{c|}{$^{13}$CCCH$^+$} & \multicolumn{2}{c|}{C$^{13}$CCH$^+$} &
\multicolumn{2}{c}{CC$^{13}$CH$^+$} \\

\hspace{-0.12in} Mode \hspace{-0.12in} & \multicolumn{1}{c|}{Description} &
Harm. & \hspace{-0.12in} Anharm. & Harm. &
\hspace{-0.12in} Anharm. & Harm. & \hspace{-0.12in} Anharm. & Harm. &
\hspace{-0.12in} Anharm. & Harm. & \hspace{-0.12in} Anharm.\\

\hline

$\nu_1(\sigma)$ & C$_1-$H stretch     & 3309.7 & 3167.8$^*$ & 2580.0 & 2502.2     & 3309.7 & 3166.7$^*$ & 3308.2 & 3163.7$^*$ & 3293.1 & 3152.0$^*$\\
$\nu_2(\sigma)$ & C$_2-$C$_3$ stretch & 2142.7 & 2096.3$^*$ & 2015.5 & 1972.2     & 2136.4 & 2090.4$^*$ & 2088.4 & 2044.9$^*$ & 2126.1 & 2080.0$^*$\\
$\nu_3(\sigma)$ & C$_1-$C$_2$ stretch & 1189.3 & 1194.1$^*$ & 1162.8 & 1162.2     & 1161.3 & 1167.2$^*$ & 1188.5 & 1191.2$^*$ & 1172.7 & 1178.5$^*$\\
$\nu_4(\pi)$ & H$-$C$_1-$C$_2$ bend   & 805.8  & 782.3      & 641.7  & 625.8$^*$  & 805.8  & 782.2      & 804.3  & 781.0      & 799.0  & 776.0 \\
$\nu_5(\pi)$ & C$_1-$C$_2-$C$_3$ bend & 124.0  & 114.2      & 117.8  & 110.6$^*$  & 123.2  & 113.4      & 121.0  & 111.2      & 123.7  & 114.1 \\

\hline\hline
\end{tabular}
$^*$Denotes states in Fermi resonance.

\end{table}
\endgroup

\end{document}